\documentclass[prc,twocolumn,floatfix,groupedaddress,nofootinbib,preprintnumbers,amsmath,amssymb,amsfonts,superscriptaddress,widetable,href]{revtex4}
\usepackage{amsmath,amssymb,amsfonts}
\usepackage{hyperref}
\usepackage{ulem}
\usepackage{graphicx}
\usepackage{color}
\usepackage{longtable}

\allowdisplaybreaks

\begin{document}

\title{Density dependence of symmetry energy and neutron skin thickness revisited using relativistic
mean field models with nonlinear couplings}
\author{C. Mondal}
\email{mondal@lpccaen.in2p3.fr}
\address{Laboratoire de Physique Corpusculaire, CNRS, ENSICAEN,
UMR6534, Université de Caen Normandie, F-14000, Caen Cedex, France}

\begin{abstract}
The correlation between neutron skin-thickness of a nucleus with neutron excess and 
density slope parameter of symmetry energy is assessed as a function of density using 
relativistic mean field models containing non-linear couplings among different mesons. 
Models with larger skin were found to probe the density slope parameter even at suprasaturation 
densities, whereas models with smaller skin were observed to be sensitive 
only at specific subsaturation density, connected to the average density 
of nuclei. Possible reasons behind this density dependence are explored 
systematically. These results might be model specific, which need to be 
reassessed in other type of interactions existing in the literature. Nevertheless, 
extrapolating the predictions at high densities from models, which are 
optimized by data at saturation or subsaturation densities, needs to be 
handled with care.

\end{abstract}

\maketitle
\section{Introduction}
Isovector part of the nuclear interaction poses some unique challenges 
in the understanding of the nucleonic systems. For example, behaviors of neutron 
rich nuclei near the drip-line change heavily from the stable nuclei; 
accurate knowledge about them might facilitate a better understanding 
of the nucleosynthesis of heavy nuclei \cite{Myers80}. The origin of neutron 
skins in nuclei with large neutron excess is also attributed to the 
isospin sensitivity of the nuclear interaction \cite{Myers69, Myers80, 
Moller12, Centelles09, Agrawal12}. Static properties of neutron stars (NS) {\it e.g.} 
mass or radius, as well as the evolution of core collapse supernovae 
are very sensitive to the neutron-proton asymmetry in the system 
\cite{Chen14, Steiner05, Janka07, Oertel17}. During the inspiral of 
a NS-NS merger, the gravitational field of one companion induces a tidal 
field on the other inducing a tidal deformation. This might be 
influenced by the neutron-proton content of the system \cite{Guven20}.
The knowledge on the possibilities of direct Urca cooling processes 
\cite{Todd-Rutel05, Lattimer91, Steiner05} hinges on the accurate 
determination of the proton fraction inside the core of a neutron 
star; this is clearly influenced by the isovector nature of the 
nuclear interaction. On these key questions, symmetry energy and 
its density derivatives can shed some light; finding precise constraints 
on them is thus imperative to have a better understanding of the 
isovector part of the nuclear force. 

The density dependence of symmetry energy is mostly governed by the density slope 
parameter $L$, which is connected to the density ($\rho$) derivative of the symmetry 
energy $C_{sym}(\rho)$ as, 
\begin{eqnarray}
L=3\rho\left (\frac{\partial C_{sym}(\rho)}{\partial\rho}\right).
\label{def_L}
\end{eqnarray}
Enormous efforts have been made in recent times 
to constrain the density slope parameter at saturation $L_0$ ($=L(\rho_0)$,
where saturation density $\rho_0\sim 0.16$ fm$^{-3}$) and subsaturation 
($0.7\rho_0$) densities \cite{Centelles09, 
Roca-Maza11, Agrawal12, Moller12, Jiang12, Zhang13, Chen14, Fattoyev14, 
Mondal15, Mondal16, Mondal16b,  Brown17} from a plethora of 
different types of experimental quantities. Concerns 
have been raised recently on constraining further the second order density dependence 
of symmetry energy, $K_{sym} \left(\sim 9\rho^2\frac{\partial^2 C_{sym}(\rho)}
{\partial\rho^2}\right)$ \cite{Mondal17, Margueron18a, Zhang18, Mondal18}, which might 
play some important roles at the core of a neutron star. The 
importance of the high density behavior of the symmetry energy in several 
astrophysical aspects is pointed out in different contemporary calculations \cite{Xie19, 
Xie20, Zhang20, Zhang21, Li21, Mondal21, Imam21}. However, the density slope 
parameter is yet to be pinned down decisively. At present, the 
accepted values of $L$ at saturation lie in between 20 - 100 MeV. This spread is rather 
large \cite{Vinas14}. Most studies on the finite nuclear properties 
as well as the astrophysical observations including the tidal 
deformability from GW170817 NS-NS merger event prefer smaller values 
of $L_0$ ($\sim 20$ - 60 MeV) \cite{Carbone10, Roca-Maza13, 
Roca-Maza13a, Mondal15, Malik18}. A larger value of neutron 
skin thickness ($\Delta r_{np}=<r_n^2>^{1/2}-<r_p^2>^{1/2}$, with 
$<r_n>^{1/2}$ and $<r_p>^{1/2}$ being the root mean square (rms) radii for the 
neutron and proton distributions, respectively) 
in $^{208}$Pb extracted by Lead (Pb) radius experiment 
(PREX-I) \cite{ Abrahamya12}, reinforced recently by PREX-II,
however, prefers larger values of $L_0\sim 80$ - 110 MeV \cite{Reed21}. 
Further refined results from PREX or upcoming Calcium radius experiment 
(CREX) \cite{JlabCREX} will shed more light in this regard. It is worthwhile to mention that 
extraction of $\Delta r_{np}$ in $^{208}$Pb from different hadronic 
probes has been made \cite{ Zenihiro10,  Krasznahorkay04,  Klos07, Friedman09}, 
although there might be some degree of model dependence involved due to 
the uncertain regime of quantum chromodynamics. 

The extraction of $L_0$ from neutron-skin thickness of a heavy nucleus 
was suggested in the early years of the present century \cite{Brown00a, Typel01, 
Horowitz01, Furnstahl02, Yoshida04, Chen05, Todd-Rutel05}. An analytic relation 
was shown by Centelles {\it et. al.} \cite{Centelles09} 
starting from Droplet model, where various symmetry energy parameters 
are connected to $\Delta r_{np}$ of a heavy nucleus. This has been 
reinforced through a wide variety of mean field models \cite{Vinas14}, 
albeit, with a mild presence of model dependence \cite{Mondal16}. 
It has been pointed out in theoretical studies \cite{Typel10, Typel14}, 
verified in an experiment very recently \cite{Tanaka21} that formation 
of $\alpha$ clusters might hinder a smooth extraction of $L$ from 
$\Delta r_{np}$ of nuclei with neutron excess. There has also been some debate 
whether the correlation between $L$ and $\Delta r_{np}$ should be at 
saturation or subsaturation densities \cite{Zhang13, Zhang14b, Zhang15}, as 
the average densities associated to nuclei are lower than the saturation 
density. A systematic study of the density dependence of this correlation 
between $L$ and $\Delta r_{np}$ in models probing different values of 
$\Delta r_{np}$ might help to elucidate more information. 
The present study is intended to do that using relativistic mean field (RMF)
models. 

The paper is organized as follows. In Sec. \ref{formalism} the RMF model 
used in the present calculation along with the fitting protocol is described. 
The results obtained in this work are described in Sec. \ref{results}. Finally, 
a brief summary is provided in Sec. \ref{summary}.

\section{Formalism}\label{formalism}
The effective RMF Lagrangian used in this work, based on usual Yukawa couplings between 
the nucleonic field $\psi$ and three mesonic fields, isoscalar-scalar 
$\sigma$, isoscalar-vector $\omega_{\mu}$ and isovector-vector 
$\boldsymbol\rho_{\mu}$ and electromagnetic field $A_{\mu}$ along with 
different couplings among the mesons, is given by \cite{Todd-Rutel05, 
Boguta77, Boguta83}
\begin{eqnarray}
\label{Lagrangian} 
{\cal L}_{\it int}=&&\overline{\psi}\left [g_{\sigma} \sigma -\gamma^{\mu} \left (g_{\omega }
\omega_{\mu}+\frac{1}{2}g_{\mathbf{\boldsymbol\rho}}\boldsymbol\tau .
{\boldsymbol\rho}_{\mu}+\frac{e}{2}(1+\tau_3)A_{\mu}\right ) \right ]\psi\nonumber
\\ &&-\frac{{\kappa_3}}{6M}
g_{\sigma}m_{\sigma}^2\sigma^3-\frac{{\kappa_4}}{24M^2}g_{\sigma}^2
m_{\sigma}^2\sigma^4+ \frac{1}{24}\zeta_0
g_{\omega}^{2}(\omega_{\mu}\omega^{\mu})^{2}\nonumber
\\ &&+\frac{\eta_{2\boldsymbol\rho}}{4M^2}g_{\omega}^2m_{\boldsymbol\rho
}^{2}\omega_{\mu}\omega^{\mu}\boldsymbol\rho_{\nu}\boldsymbol\rho^{\nu}.  
\end{eqnarray}
$m_{\sigma}$, $m_{\omega}$ and $m_{\boldsymbol\rho}$ are the masses of the 
$\sigma,\ \omega$ and $\boldsymbol\rho$ mesons, respectively. The values of 
$m_{\omega} = 782.5$ MeV and $m_{\boldsymbol\rho} = 763$ MeV are kept fixed 
in this calculation. The values of $m_{\sigma}$ along with other coupling 
constants in Eq. (\ref{Lagrangian}) {\it e.g.} $g_{\sigma}, g_{\omega }, 
g_{\mathbf{\boldsymbol\rho}}, \kappa_3, \kappa_4, \zeta_0$ and $ 
\eta_{2\boldsymbol\rho}$ are fitted to experimental data optimizing an 
objective function. The cross-coupling between the $\omega_{\mu}$ and 
$\boldsymbol\rho_{\mu}$ mesons, with coupling constant $\eta_{2
\boldsymbol\rho}$, can facilitate one to obtain models with different 
density dependence of symmetry energy (and eventually different $\Delta r_{np}$), 
but keeping the isoscalar behavior very similar \cite{Furnstahl02, Sil05}. 

Four different models were obtained, namely, MOD16, MOD18, MOD19 and MOD23 
by fitting binding energies $BE$ and 
charge radii $r_{ch}$ of 12 closed-shell spherical nuclei across the whole nuclear 
chart, along with the value of $\Delta r_{np}$ 
in $^{208}$Pb fixed to four different values {\it viz.} 0.16, 0.18, 0.19 
and 0.23 fm, respectively. 
The goal of this systematic way of fixing $\Delta r_{np}$ in $^{208}$Pb is 
to study their effects in different symmetry energy parameters as a function 
of density, in particular, how it affects the extraction of the $L$ parameter 
from the correlation between $L$ and $\Delta r_{np}$ in $^{208}$Pb. 
The objective function minimized to optimize the 
different coupling constants in Eq. (\ref{Lagrangian}) is given by, 
\begin{equation}
\chi^2({\bf p}) =\frac{1}{N_d - N_p}\sum_{i=1}^{N_d} \left (\frac{ \mathcal{O}_i^{exp}
- \mathcal{O}_i^{th}({\bf p})}{\Delta\mathcal{O}_i}\right )^2.
\label {chi2}
\end{equation}
The number of experimental data points is 
$N_d$ and the number of fitted parameters is $N_p$.  The experimental and the
corresponding theoretically obtained values of an observable are 
$\mathcal{O}_i^{exp}$ and $\mathcal{O}_i^{th}({\bf p})$, respectively.
$\Delta\mathcal{O}_i$ is the adopted error of an observable.  The minimum
value of the objective function $\chi_0^2$ corresponds to the 
$\chi^2$ at ${\bf p_0}$; ${\bf p_0}$ being the optimal values of
the parameters. Once the objective function is minimized, one can calculate 
 the covariance between two quantities $\mathcal{A}$ and $\mathcal{B}$ (or 
 variance putting $\mathcal{A}=\mathcal{B}$) as \cite{Brandt97},
\begin{equation}
\overline{\Delta \mathcal{A}\Delta\mathcal{B}}=\sum_{\alpha\beta}\left(
\frac{\partial \mathcal{A}}{\partial
\rm{p}_{\alpha}}\right)_{\bf p_0}\mathcal{C}_{\alpha\beta}^{-1}\left(\frac
{\partial \mathcal{B}}{\partial \rm{p}_{\beta}}
\right)_{\bf p_0}.
\label{deltaab}
\end{equation}
The correlation coefficient between $\mathcal{A}$ and $\mathcal{B}$ is specified by
\begin{equation}
  {c}_\mathcal{AB} =
  \frac{\overline{\Delta \mathcal{A}\,\Delta \mathcal{B}}}
       {\sqrt{\overline{\Delta \mathcal{A}^2}\;\overline{\Delta \mathcal{B}^2}}}.
\label{corr}
\end{equation}
One should note that $\mathcal{A}$ and $\mathcal{B}$ can be observables as well as parameters. 
An inverted element of the curvature matrix $\mathcal{C}_{\alpha\beta}^{-1}$ appearing in 
Eq. (\ref{deltaab}) is connected to the $\chi^2$ as,
\begin{equation}
 \mathcal{C}_{\alpha\beta}=\frac{1}{2}\Big(\frac{\partial^2 \chi^2(\mathbf{p})}
{\partial {\rm {p}_{\alpha}}\partial {\rm{p}_{\beta}}}\Big)
_{\mathbf{p}_0}.
\label{Cmatrix}
\end{equation}

In Table \ref{data}, different observables $\mathcal{O}$ fitted 
in the present work, their experimental values \cite{Wang12,Angeli13}, 
adopted errors on them ($\Delta\mathcal{O}$) \cite{Klupfel09},
along with their fitted values are provided. The $\chi_0^2$ values 
for all the four models obtained here are very similar, 
lying in between 0.7 and 0.8, which ensures a good quality of the 
fit for all the four models.
\begin{table}[h]
\caption{\label{data}
Observables $\mathcal{O}$, adopted errors on them
$\Delta\mathcal{O}$, corresponding experimental data (Expt.) \cite{Wang12,Angeli13} and their
best-fit values for MOD16, MOD18, MOD19 and MOD23. $BE$ and $r_{ch}$ corresponds to
binding energy and charge radius of a nucleus, respectively and $\Delta r_{np}$ is
the neutron-skin thickness of the corresponding nucleus. Values of $BE$ are given in
units of MeV and $r_{ch}$ and $\Delta r_{np}$ in fm. }
 \begin{ruledtabular}
\begin{tabular}{cccccccc}
& $\mathcal{O}$ &$\Delta\mathcal{O}$ & Expt. &
	MOD16 & MOD18 & MOD19 & MOD23\\
\hline
	$^{16}$O & $BE$ &4.0 & 127.62 & 128.22 & 128.12 & 127.99 & 127.80 \\
	& $r_{ch}$ &0.04 & 2.699 & 2.693 & 2.697 & 2.700 & 2.703 \\
	$^{40}$Ca& $BE$ &3.0 & 342.05 & 343.41 & 343.47 & 343.31 & 343.27 \\
	& $r_{ch}$ &0.02 & 3.478 & 3.448 & 3.452 & 3.455 & 3.458 \\
	$^{48}$Ca& $BE$ &1.0 & 416.00 & 415.08 & 415.12 & 415.09 & 415.18 \\
	& $r_{ch}$ &0.04 & 3.477 & 3.437 & 3.437 & 3.437 & 3.437 \\
	$^{56}$Ni& $BE$ &5.0 & 483.99 & 484.42 & 484.13 & 484.10 & 483.37 \\
	& $r_{ch}$ &0.18 & 3.750 & 3.682 & 3.686 & 3.688 & 3.699 \\
	$^{68}$Ni& $BE$ &2.0 & 590.41 & 592.42 & 592.65 & 592.64 & 592.81 \\
	$^{90}$Zr& $BE$ &1.0 & 783.90 & 783.19 & 783.16 & 783.08 & 783.09 \\
	& $r_{ch}$ &0.02 & 4.269 & 4.263 & 4.263 & 4.264 & 4.265 \\
	$^{100}$Sn& $BE$ &2.0 & 825.30 & 828.00 & 828.00 & 828.12 & 828.06 \\
	$^{116}$Sn& $BE$ &2.0 & 988.68 & 986.97 & 987.13 & 987.33 & 987.55 \\
	& $r_{ch}$&0.18 & 4.625 & 4.622 & 4.621 & 4.621 & 4.620 \\
	$^{132}$Sn& $BE$ &1.0 & 1102.84 & 1102.87 & 1102.88 & 1102.95 & 1103.07 \\
	& $r_{ch}$&0.02 & 4.709 & 4.713 & 4.710 & 4.709 & 4.707 \\
	$^{144}$Sm& $BE$ &2.0 & 1195.73 & 1195.90 & 1195.86 & 1195.81 & 1196.12 \\
	& $r_{ch}$&0.02 & 4.952 & 4.953 & 4.954 & 4.954 & 4.955 \\
	$^{208}$Pb& $BE$ &1.0 & 1636.43 & 1636.64 & 1636.59 & 1636.55 & 1636.42 \\
	& $r_{ch}$&0.02 & 5.501 & 5.531 & 5.529 & 5.529 & 5.528 \\
	& $\Delta r_{np}$&0.01 & - & 0.163 & 0.182 & 0.192 & 0.230 \\
\end{tabular}
\end{ruledtabular}
\end{table}

With the optimized set of parameters, symmetry energy as a function of 
density can be calculated using the Lagrangian described in Eq. (\ref{Lagrangian})
as
\begin{eqnarray}
\label{csymrho}
C_{sym}(\rho)&=&\frac{k_F^2}{6(k_F^2+{M^*}^2)^{1/2}}+\frac{g_{\boldsymbol
\rho}^2}{12\pi^2}\frac{k_F^3}{{m_{\boldsymbol\rho}^*}^2},\\
\label{mstar}
\text{with,}\ M^*&=&M-g_{\sigma} \sigma,\\
\label{mrhostar}
\text{and},\ {m_{\boldsymbol\rho}^*}^2 &=& m_{\boldsymbol\rho}^2\left(1+\frac{1}{2M^2}
\eta_{2\boldsymbol\rho}g_{\omega}^2\omega^2\right).
\end{eqnarray}
Here, $k_F$ is nucleon Fermi-momentum connected to the density as, $k_F=
\left(\frac{3\pi^2\rho}{2}\right)^{1/3}$; $M^*$ and $M$ are Dirac effective 
mass and free mass of nucleon, respectively; $m_{\boldsymbol\rho}^*$ is 
the effective mass of $\boldsymbol\rho$ meson. The kinetic part of the 
symmetry energy $C_{sym}$ in Eq. (\ref{csymrho}) depends on the Dirac effective 
mass, which depends on the $\sigma$ field along with the coupling constant 
$g_{\sigma}$. The interaction part depends on the isovector coupling 
constants $g_{\boldsymbol\rho}$ and $\eta_{2\boldsymbol\rho}$ through the 
effective mass of $\boldsymbol\rho$ meson in Eq. (\ref{mrhostar}).

\section{Results and Discussions}\label{results}
\begin{figure}[]{}
{\includegraphics[height=3.5in,width=3.2in,angle=-90]{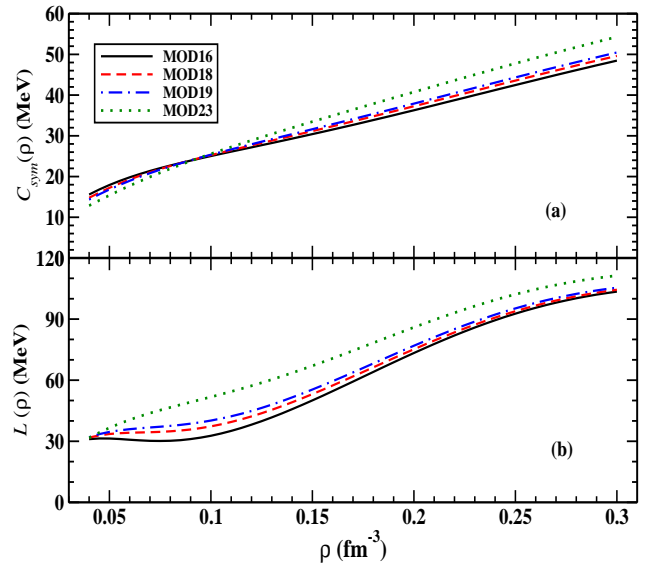}}
\caption{\label{ljrho}
(Color online) Symmetry energy $C_{sym}$ and density slope parameter $L$ plotted 
	as a function of density $\rho$ obtained with MOD16, MOD18, MOD19 and 
	MOD23.}
\end{figure}
In Fig. \ref{ljrho}, symmetry energy $C_{sym}$ (Fig. \ref{ljrho}(a)) and 
density slope parameter $L$ (Fig. \ref{ljrho}(b)) as a function of density are 
plotted for the four models MOD16, MOD18, MOD19 and MOD23 as discussed before. 
The behavior of $C_{sym}$ varies slightly going from MOD16 to MOD23, becoming 
slightly stiffer in the latter. All the four models in discussion have a
crossover at around $\rho\simeq 0.1$ fm$^{-3}$, with $C_{sym}(\rho=0.1) 
\approx 24$ MeV. This is not surprising keeping in mind the fact that the finite 
nuclei constrain the symmetry energy at a subsaturation density $\sim 0.7 
\rho_0$ with its value $\sim 24$ MeV \cite{Trippa08}. In contrast, density slope 
parameter $L$ varies the most around this subsaturation density for the models 
considered. For MOD16, $L$ has a decreasing trend till $\rho\sim 0.1$ 
fm$^{-3}$ after which it keeps increasing. On the other extreme MOD23 shows 
almost a monotonous increasing trend in $L$ with density.

\begin{figure}[]{}
{\includegraphics[height=3.5in,width=3.2in,angle=-90]{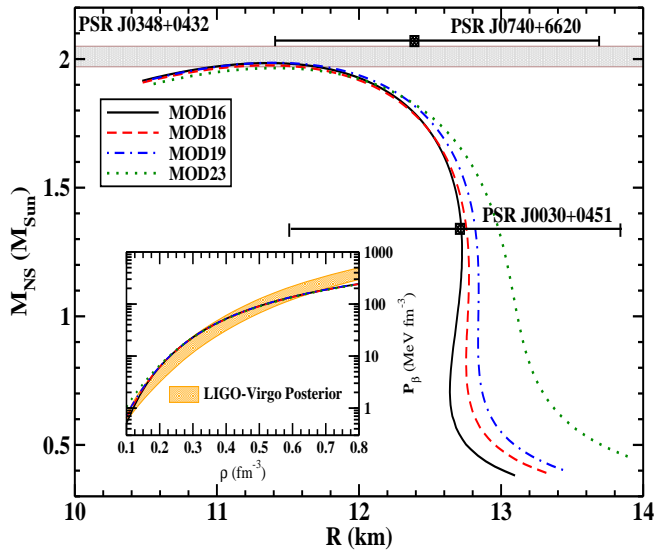}}
\caption{\label{mr}
(Color online) Mass-radius relation for neutron stars obtained with 
	MOD16, MOD18, MOD19 and MOD23. The constraints on the mass for 
	pulsar PSR J0348+0432 ($=2.01\pm0.04 M_{\odot}$) obtained 
	from radio astronomy observation \cite{Antoniadis13} is given as 
	the grey shaded region. Bounds on the radii from the observations  of pulsars 
	PSR J0030+0451 and PSR J0740+6620 by NICER \cite{Riley21, Miller21} 
	collaboration are also indicated. (Inset): The pressure on the beta-equilibrated 
	matter from the four models along with the posterior band from LIGO-Virgo 
	\cite{Abbott18} collaboration.}
\end{figure}
In Fig. \ref{mr}, mass-radius relation of a neutron star, calculated by solving 
the Tolman-Oppenheimer-Volkoff (TOV) equations using the Lagrangian in Eq. 
(\ref{Lagrangian}), is plotted for 
all the four models obtained in this work (see Table \ref{data}). The 
equation of state (EoS) of NS gets stiffer with increasing value of $L(\rho_0)$ \cite{Cavagnoli11}. 
One can observe an overall increase in the radii obtained in the models 
with bigger $\Delta r_{np}$ in $^{208}$Pb, as $\Delta r_{np}$ and $L
(\rho_0)$ are linearly correlated to each other. For reference, the 
constraint on the mass of pulsar PSR J0348+0432 ($=2.01\pm0.04 
M_{\odot}$) obtained from radio astronomy using the Shapiro delay 
technique \cite{Antoniadis13} is provided in Fig. \ref{mr}. All the models marginally 
satisfy the constraint provided by this observation. Overall, pretty 
similar features are observed in $M$-$R$ plane for all the fours models 
obtained in the present analysis, especially the heavier NSs have almost 
same radii in each model. Constraints on the radii of pulsars PSR J0030+0451 
and PSR J0740+6620 obtained from NICER collaboration \cite{Riley21, Miller21} 
are also provided for comparison. The constraint on the heavier star is fairly away 
from the models considered, however, if the two dimensional joint posterior 
for the mass-radius is taken into account, there might be a little overlap 
among them \cite{Dinh-Thi21}. In the inset of Fig. \ref{mr}, the pressure of 
beta equilibrated matter $P_{\beta}(\rho)$ is plotted as a function of density 
for the four models obtained in this work, simultaneously with the posterior 
predicted by the LIGO-Virgo collaboration \cite{Abbott18}. All the models overlap 
quite consistently with the LIGO-Virgo posterior throughout the concerned density range. 
It is worthwhile to mention here that for all the four models used in this 
work, the BPS crust EoS \cite{Baym71} is used below the fixed transition density 
0.08 fm$^{-3}$. A more sophisticated way to deal with the crust-core transition 
would change the radii of small mass NSs at most by 100 - 200 meters, but it would not 
change the conclusions made above.

\begin{figure}[h]{}
{\includegraphics[height=3.5in,width=3.2in,angle=-90]{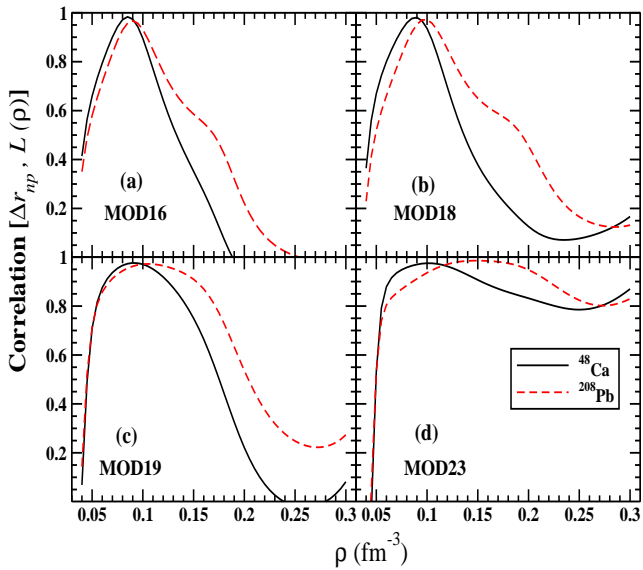}}
\caption{\label{skin_l_rho}
(Color online) Correlation coefficient between neutron skin $\Delta r_{np}$ 
	in $^{48}$Ca and $^{208}$Pb with density slope parameter $L$ is plotted as 
	a function of density $\rho$ for MOD16, MOD18, MOD19 and MOD23.}
\end{figure}
In Fig. \ref{skin_l_rho}, correlation (Eq. (\ref{corr})) between 
$\Delta r_{np}$ of nuclei and density slope parameter $L$ is plotted as a 
function of density for $^{48}$Ca and $^{208}$Pb obtained with MOD16, 
MOD18, MOD19 and MOD23. The results for $^{48}$Ca and $^{208}$Pb are 
found to be very similar. This is not surprising keeping in mind the fact that the 
neutron skins in nuclei are quite well correlated among themselves in mean-field 
models \cite{Mondal16b}. The correlation coefficient has a very 
prominent peak, both for $^{48}$Ca and $^{208}$Pb in the models with 
lower $\Delta r_{np}$ (MOD16 or MOD18), which broadens in MOD19 and 
becomes very wide, appearing almost flat throughout the concerned density range in 
MOD23. The maximum correlation coefficient between $\Delta r_{np}$ of 
$^{208}$Pb and $L$ parameter appears at densities 0.09, 0.1, 0.105 and 0.15 
fm$^{-3}$ with correlation coefficients 0.967, 0.972, 0.972 and 0.986 
for the models MOD16, MOD18, MOD19 and MOD23, respectively. The results 
for $^{48}$Ca and $^{208}$Pb are quite similar for all the four 
models as the correlation coefficient between their corresponding $\Delta r_{np}$'s 
are quite high with the minimum being for MOD19  $\sim -0.79$ and maximum for MOD18 $\sim -0.92$. 
One can also observe that the peaks for $^{208}$Pb are at 
slightly higher densities for all the cases in comparison to those 
corresponding to $^{48}$Ca. This shows that the average 
density associated with $^{208}$Pb is slightly 
larger than that in $^{48}$Ca. The density dependence of this 
correlation has some consequences at high densities. It seems that 
if an RMF model with non-linear couplings among different mesons 
as in Eq. (\ref{Lagrangian}) probes a higher value of $\Delta r_{np}$ 
in $^{208}$Pb, it probes quite unambiguously the density slope parameter even at 
suprasaturation densities. Incidentally, the models 
SINPB and SINPA obtained in a previous work \cite{Mondal16} 
extract very similar $\Delta r_{np}$ in $^{208}$Pb as those obtained 
from MOD23 and MOD18, respectively. There, the neutron skins were 
not part of the fit data. However, the density dependence of the 
correlation between $\Delta r_{np}$ and $L$ shows almost the same 
behavior. To understand this in a systematic way, the following 
analysis is done.

\begin{figure}[h]{}
{\includegraphics[height=1.75in,width=3.5in]{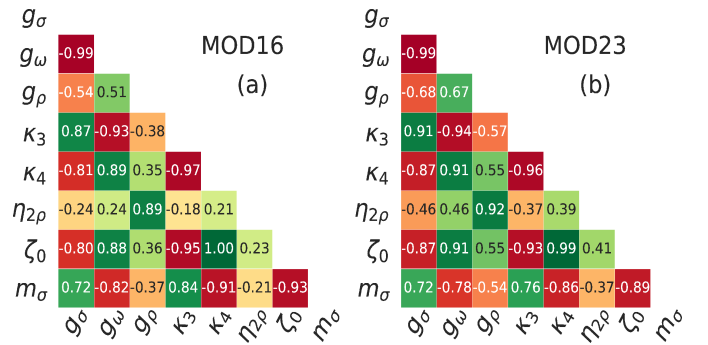}}
\caption{\label{corr_16_23}
	(Color online) Correlation coefficients obtained with Eq. (\ref{corr}),  
	among different model parameters present in Eq. (\ref{Lagrangian}) for 
	MOD16 and MOD23 are displayed in panel (a) and (b), respectively.}
\end{figure}
Correlation between two observables calculated from a model depends 
in an involved way on the interdependence of different model parameters 
among themselves. The correlation coefficients among different 
parameters in Eq. (\ref{Lagrangian}) are plotted in Fig. \ref{corr_16_23} for 
MOD16 and MOD23, which represent the two ends of the $\Delta r_{np}$ 
probed by the models obtained in the present analysis. There are many 
strong correlations present among the parameters. The parameters 
$g_{\boldsymbol\rho}$ and $\eta_{2\boldsymbol\rho}$, which primarily 
control the isovector part of the interaction, are quite independent 
from the rest of the parameters ({\it i.e.} correlation coefficients 
are small); $g_{\boldsymbol\rho}$ has slightly more 
dependence on other parameters than $\eta_{2\boldsymbol\rho}$. 
\begin{figure}[h]{}
{\includegraphics[height=3.5in,width=3.2in,angle=-90]{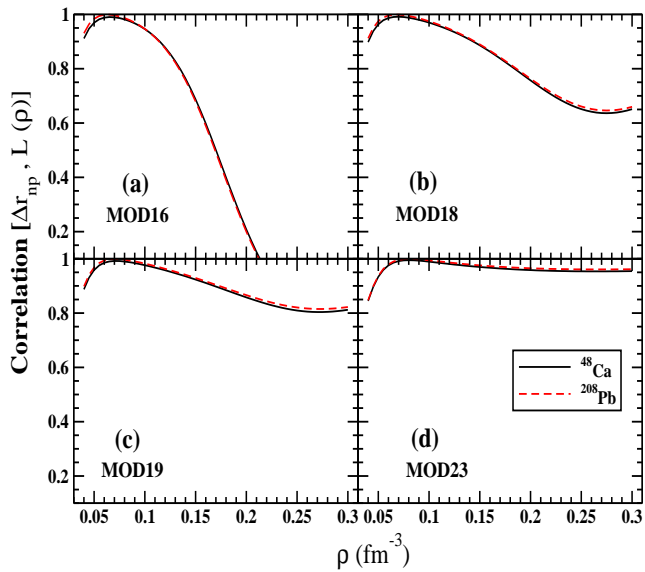}}
\caption{\label{off_diagonal}
	(Color online) Same as Fig. \ref{skin_l_rho} ignoring the off-diagonal 
	elements present in the correlation matrix for the model parameters which 
	are displayed as in Fig. \ref{corr_16_23}.}
\end{figure}
However, they are highly correlated to one another. This complex 
interrelation among the parameters clouds the clear knowledge of the 
correlation among two observables calculated from these models. This 
is why understanding the density dependence of the correlation between 
$\Delta r_{np}$ and $L$ in Fig. \ref{skin_l_rho} is not very 
straightforward starting from the correlation among parameters 
displayed in Fig. \ref{corr_16_23}.

To escape the complex interdependences existing in the multi-dimensional 
parameter space, as a simplistic assumption, all the inter-correlations 
existing in Fig. \ref{corr_16_23} are ignored and the study of Fig. 
\ref{skin_l_rho} is repeated and presented in Fig. \ref{off_diagonal}. The general 
behavior of the correlation between $\Delta r_{np}$ and $L$ as a 
function of density remains almost unaltered. In models with smaller 
$\Delta r_{np}$ for $^{208}$Pb, {\it i.e.} for MOD16 or MOD18, the 
correlation peaks at a subsaturation density and rapidly decreases 
at higher densities. The rate of the decrease diminishes when one 
enters in models with larger $\Delta r_{np}$. The correlation 
coefficients governed by Eqs. (\ref{deltaab},\ref{corr}) are now 
totally dependent on the Jacobians $\left(=\left(\frac{\partial 
\mathcal{A}}{\partial p_{\alpha}}\right)_{p_0}\right)$ present in 
Eq. (\ref{deltaab}).

\begin{figure}[t]{}
{\includegraphics[height=3.5in,width=3.2in,angle=-90]{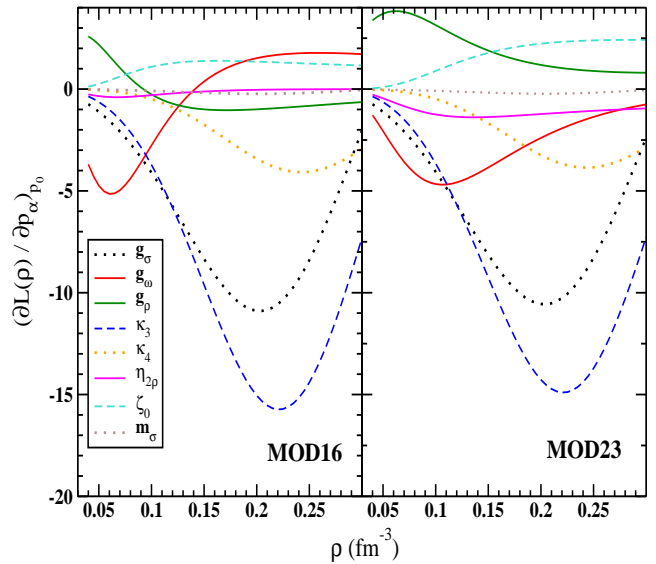}}
\caption{\label{jacobian_rho}
(Color online) Density dependence of the Jacobians involving density slope 
	parameter $L$ and different model parameters of Eq. (\ref{Lagrangian}) 
	are presented for MOD16 and MOD23 in panels (a) and (b), respectively 
	(see text for details).}
\end{figure}
In Fig. \ref{jacobian_rho} Jacobians for the density slope parameter $L$ 
({\it i.e} $\mathcal{A}=L$) is plotted as a function of density for 
MOD16 and MOD23. As the variation of the $\Delta r_{np}$ with respect 
to the parameters does not change with density, the variation in the 
correlation pattern observed in Fig. \ref{off_diagonal} (consequently 
Fig. \ref{skin_l_rho}) is essentially determined by the change in 
$\left(\frac{\partial L(\rho)}{\partial p_{\alpha}}\right)_{p_0}$ 
as a function of density. Also, the Jacobians for $\Delta r_{np}$ 
involving all the parameters are found to be very similar for 
MOD16 and MOD23. It can be seen in Fig. \ref{jacobian_rho} that the 
primary differences in the density dependences of Jacobians exist 
corresponding to the parameters $g_{\omega}, g_{\boldsymbol\rho}$ 
and $\eta_{2\boldsymbol\rho}$. The differences in the behavior of Jacobians for 
density slope parameter $L$ involving $g_{\boldsymbol\rho}$ and $\eta_{2\boldsymbol\rho}$ is 
easy to understand, as they control the isovector part of the 
interaction (cf. Eqs. (\ref{csymrho},\ref{mrhostar})). The dissimilarity involving 
$g_{\omega}$ can be realized by looking at the correlation between 
$g_{\boldsymbol\rho}$ and $g_{\omega}$ in Fig. \ref{corr_16_23} 
which transforms from a weak correlation (correlation coefficient 
0.51) in MOD16 to a moderate correlation (with coefficient 0.67) 
in MOD23.

\section{Summary}\label{summary}
To summarize, the density dependence of the correlation between 
neutron skin-thickness and the density slope parameter is systematically 
studied to shed some light on the nature of the isovector nuclear 
interaction. Relativistic models of FSU type \cite{Todd-Rutel05} containing 
the cross coupling between $\omega$ and $\boldsymbol\rho$ mesons 
are used. Models with a wide range of input neutron skin thickness 
in $^{208}$Pb as pseudo-data, along with experimental data on binding 
energies and charge radii of nuclei across the whole nuclear chart 
are used. It was observed that the behavior 
of the symmetry energy and density slope parameter as a function of density 
or the mass-radius relation in neutron stars is only marginally 
different for these different models, so also the relation of 
pressure in beta-equilibrated matter to density. However, a sharp contrast 
was found in the density dependence of the correlation between 
neutron skin in nuclei and the density slope parameter. 
A model probing a smaller value of neutron skin, shows a peak 
in the aforementioned correlation at a subsaturation density 
which progressively flattens and shifts to a high value of density 
in a model corresponding to a high value of neutron 
skin. To understand this contrasting behavior, correlation systematics among the model 
parameters were studied. In a follow up simplistic analysis it was further shown that 
if the correlation among the model parameters are ignored 
{\it a priori}, the overall behavior of the correlation pattern 
does not change as such. This discloses that the isovector 
part of the interaction changes its density dependence in models 
with different values of neutron skin. This might be a specific 
feature of the model used in the present analysis (Eq. 
(\ref{Lagrangian})). Nevertheless, these kind of features should as well be 
explored in other type of interactions, both relativistic and 
non-relativistic, as extrapolations of models are often employed 
to study high density behaviors of nuclear interactions which are 
primarily obtained by fitting data pertaining to saturation or 
subsaturation densities.

\section{Acknowledgements} 
The author acknowledges 
Jadunath De and Gagandeep Singh for carefully reading the manuscript and 
their useful suggestions. This work was partially supported by
the IN2P3 Master Project “NewMAC”.


\end{document}